\documentclass[aps,twocolumn]{revtex4-1}
\usepackage{amsmath}
\usepackage{color}
\usepackage{floatrow}
\usepackage{amssymb}
\usepackage{graphicx}
\usepackage{blindtext}
\usepackage{multirow}
\usepackage{booktabs}
\usepackage{chngcntr}
\usepackage{subfigure}
\usepackage[table,xcdraw]{xcolor}
\usepackage{booktabs}
\usepackage{hhline}
\usepackage{float}    

\def\CrO{CrO$_2$}
\def\MnO{MnO$_2$}

\def\etal{{\it et al. }}

	\def\etal{{\it et al. }}

\begin{document}

\title{Porous CrO$_2$: a ferromagnetic half-metallic member in sparse hollandite oxide family}
\author{Sujoy Datta}
\email{sujoydatta13@gmail.com, sujoy.datta@utoronto.ca}
\affiliation{Department of Physics, University of Toronto,	60 Saint George Street, Toronto, Ontario M5S 1A7, Canada}
\affiliation{Kadihati KNM High School, Ganti, Kolkata 700135, India} \thanks{present address.}

\begin{abstract}
	A stable polymorph of CrO$_2$ is predicted using PBE+U method. The porous material is isostructural with $\alpha-$MnO$_2$ making it the second transition metal oxide in sparse hollandite group of materials. However, unlike the anti-ferromagnetic semiconducting character of the $\alpha-$MnO$_2$, it is found to be a ferromagnetic half-metal. At Fermi level, the hole pocket has ample contribution from O$-2p$ orbital, though, the electron pocket is mostly contributed by Cr$-3d_{xy}$ and Cr$-3d_{x^2-y^2}$. A combination of negative charge transfer through orbital mixing and extended anti-bonding state near Fermi level is responsible for the half-metallic ferromagnetic character of the structure. A comparative study of rutile and hollandite CrO$_2$ and hollandite MnO$_2$ structures delineate the interplay between structural, electronic and magnetic properties. The material shows a robust magnetic character under hydrothermal pressure, as well as, the band topology is conserved under uniaxial strain.  Moderate magneto-crystalline anisotropy is observed and it shows a correspondence with the anisotropy of elastic constants. Occurrence of type$-$II Weyl nodes and their evolution under pressure is explored.
\end{abstract}

\date{\today}
\maketitle

\section{Introduction}
{\par}In the field of material science, studies on transition metal oxides (TMO) is prolific. However, TMOs never get tired to amaze the scholars through their versatile character. Not only that, even the burgeoning development in first-principles electronic-structure schemes faces the challenge of describing the underlying Physics of TMOs. Beside dynamical mean free theory (DMFT), inclusion of Hubbard parameters is established as a successful approach in describing TMOs.

{\par} Though TMOs are found to form solid-state structure of various symmetries, some structures are known for their uniqueness. The porous layered hollandite structure of $\alpha-$MnO$_2$ is such an example \cite{bystrom1950}. The $2\times2$ tunnel within this structure can accommodate additional atomic species (e.g., Pb$^{ 2+}$ , B$^{2+}$, K$^+$ , etc.) while producing \cite{luo2010,li2007}. Not only it is beneficial for ion storage application for battery industry, but also such additional species can tune the electronic and magnetic properties as well \cite{gao2020,kim2020,hu2015,tseng2015,tripkovic2018}. A good number of publications in this field is the testimonial to the importance of study on this structure \cite{cockayne2012,tseng2015,wang2009,wang2019,zhang2012,mondal2020,sivakumar2021,dong2020}. However, till date, there is no report on any other hollandite TMO exhibiting similar structure with hollow tunnel except $\alpha-$MnO$_2$.

{\par} Beside Manganese, hollandite oxide structures of Vanadium, Chromium, Ruthenium and Titanium were theoretically or experimentally studied where different cations occupy the tunnel space \cite{abriel1979,carter2002,lebedev2017,stimpson2018,endo1976}. For example, in 
K$_2$V$_8$O$_{16}$ Potassium ions reside at the tunnel of the hollandite-type structure exhibiting anti-ferromagnetic (AFM) insulating ground state \cite{okada1978,isobe2006}.  In contrast K$_2$Cr$_8$O$_{16}$ has ferromagnetic (FM) insulating ground state ($T\langle 95K$) and, in the temperature range $95K \langle T \langle 180K$ it is found in metallic FM state \cite{hasegawa2009,tamada1996,mahadevan2010,toriyama2011}.

{\par} In K$_2$Cr$_8$O$_{16}$, there are two Cr$^{3+}$ and six Cr$^{4+}$ per formula unit. As Cr$^{3+}$ is more stable ionic state found in several compounds, the extraction of K$^+$ ion is a challenging task. Using electrochemical oxidation or soft-chemistry reactions, partial de-insertion of K$^+$ have been reported. This may provide a clue to mitigate the hindrance in synthesising porous hollandite \CrO~ \cite{pirrotta2012}.

{\par}In transition metal family, Chromium is one of its kind. Even at room temperature, is found to be antiferromagnetic, making it the only single elemental AFM solid \cite{marcus1998}. It shows a valency in the range of $[+1 ~to~ +6]$ and oxidize to form CrO, Cr$_2$O$_3$, CrO$_2$ and CrO$_5$. Experimentally, two phases of Chromium dioxides have been characterised, the rutile type (namely $\alpha-$\CrO, or, r$-$\CrO) and orthogonal CaCl$_2$ type (namely $\beta-$\CrO, or, o$-$\CrO). There is a second order phase transition observed from rutile to CaCl$_2$ type phase at $12-17~GPa$ \cite{maddox2006, kuznetsov2006}. Some other dynamically stable phases have also been proposed theoretically though hollandite structure have not been predicted yet \cite{bendaoud2019,kim2012,huang2018}.

{\par}Transition metals are characterised by the electrons in their $d-$orbital. In its $4+$ valance state, Cr ion has two $3d$ electrons and a vacant $4s$ shell. These two $3d$ electrons should reside at two $t_{2g}$ orbitals. Strong on-site correlation between these two electrons should result in Mott type insulating nature, however, this is not the case. Rutile type Chromium dioxide is a ferromagnetic half-metal at its ground state \cite{schwarz1986}. An explanation of the half-metallic  ferromagnetic character has been provided in terms of the double-exchange model \cite{korotin1998,katsnelson2008,kulatov1990}. This makes r$-$\CrO~ a negative charge transfer insulating materials in the Zaanen-Sawatzky-Allen (ZSA) scheme \cite{korotin1998, zaanen1985}. The O$-2p$ bands crossing the Fermi energy works as a charge (electron/hole) centre to nullify the strong electron-electron correlation between Cr$-3d$ electrons through the hybridisation. Recently the topological character of rutile and orthorhombic type \CrO ~are investigated. It is shown that type-I and type-II Weyl fermions, can emerge in these phases of chromium dioxide \cite{wang2018}.

{\par} With the technological advancement, energy storage is a bottleneck yet to be cleaned. Holey structures are efficacious candidate in storing Lithium, Sodium, Zinc or other ions prospective to battery material \cite{tompsett2013,li2012}. Both Chromium and oxygen are abundant in nature, so, h$-$\CrO~ can be a good alternative in this field. Furthermore, half-metallic ferromagnets are playing a critical role in modern spintronics devices; from magnetic sensors, spin valves to computer hardware components such as magneto-resistive random-access memory (MRAM), read head of magnetic hard drive, etc. \cite{attema2005,irkhin1994,yuasa2018,keen2002}.

{\par} In view of the rare ferromagnetic nature of CrO$_2$ in the TMO family, an investigation on the structures and their local bonding characteristics is indispensable. The aim of this article is to use the latest reliable theoretical approaches to demonstrate the mechanical, electronic and magnetic character of the proposed hollandite polymorph. To analyse and interpret the complicated electronic structures of TMOs, introduction of Hubbard term for both of on-site and off-site electron-electron interaction proved to be an efficacious tool \cite{bendaoud2019,himmetoglu2014}.

{\par} A detailed side by side investigation of the three materials, already synthesised r$-$\CrO, h$-$MnO$_2$, and the proposed h$-$\CrO~ can bridge the gap of theoretical understanding on such materials having structural intricacies as well as rich electronic and magnetic properties. Such trustworthy theoretical predictions will facilitate experimentalists a better handle on choosing their materials for a desired application, out of a plethora of possibilities.

\section{Computational Details}
{\par} We have used the Vienna Ab-initio Simulation Package (VASP) package for density functional theoretical (DFT) calculations \cite{vasp}. Projected augmented wave (PAW) basis enabled pseudo-potentials using Perdew-Burke-Ernzerhof (PBE) and PBE for solids (PBEsol) generalised gradient approximated (GGA) exchange-correlation (xc) functional have been used \cite{PBE,PBEsol}. The optimised structures are found through ionic and volume relaxations using GGA and GGA+U methods starting from different magnetic configurations. We have set the threshold of maximum force as $10^{-5}~Ry./atom$ and pressure threshold as $10^{-5}~Kbar/cell$. The convergence criteria for energy and charge densities has been set as $10^{-8}~Ry$. We have chosen kinetic-energy cut-off $520~Ry$. Fine reciprocal-space grids of dimensions $8\times8\times8$, $5\times5\times8$, $5\times5\times8$ are used for holandite, rutile and orthorhombic structures, respectively.

{\par}The electron configurations of Cr and O have been taken as [Ar]$4s^2 3d^4$ and [He]$2s^2 2p^4$. As $3d$ electrons of transition metals correlate more strongly than the limit of GGA, for proper description of strong electron-electron correlation, Hubbard U and Hund J terms have been used. While U provides intra-orbital Coulomb interaction, $U-J$ takes care of the inter=orbital Coulomb interaction between electrons.

{\par} Over the years, several non-empirical approaches have been proposed to estimate Hubbard parameters, such as constrained DFT, constrained random phase approximation (cDFT/cRPA) and linear-response formulation \cite{csacsiouglu2011,pickett1998}. These terms for r$-$\CrO~ have been calculated by constrained screening method by Korotin \etal \cite{korotin1998}. We have used these values for \CrO: $U=3.0$ and $J=0.87$.
For MnO$_2$,  $U=5.87$ is used \cite{paul2023}.

{\par} The Vesta package has been utilised to simulate X-ray diffraction (XRD) pattern for Cu$_\alpha$ radiation \cite{vesta}. For pre and post-processing Vaspkit and ElATools are utilised \cite{elatools,vaspkit}. The chemical bonding analysis has been done using Lobster package for PAW \cite{nelson2020}. 

\begin{figure}[]
	\centering	
	\includegraphics[width=\textwidth]{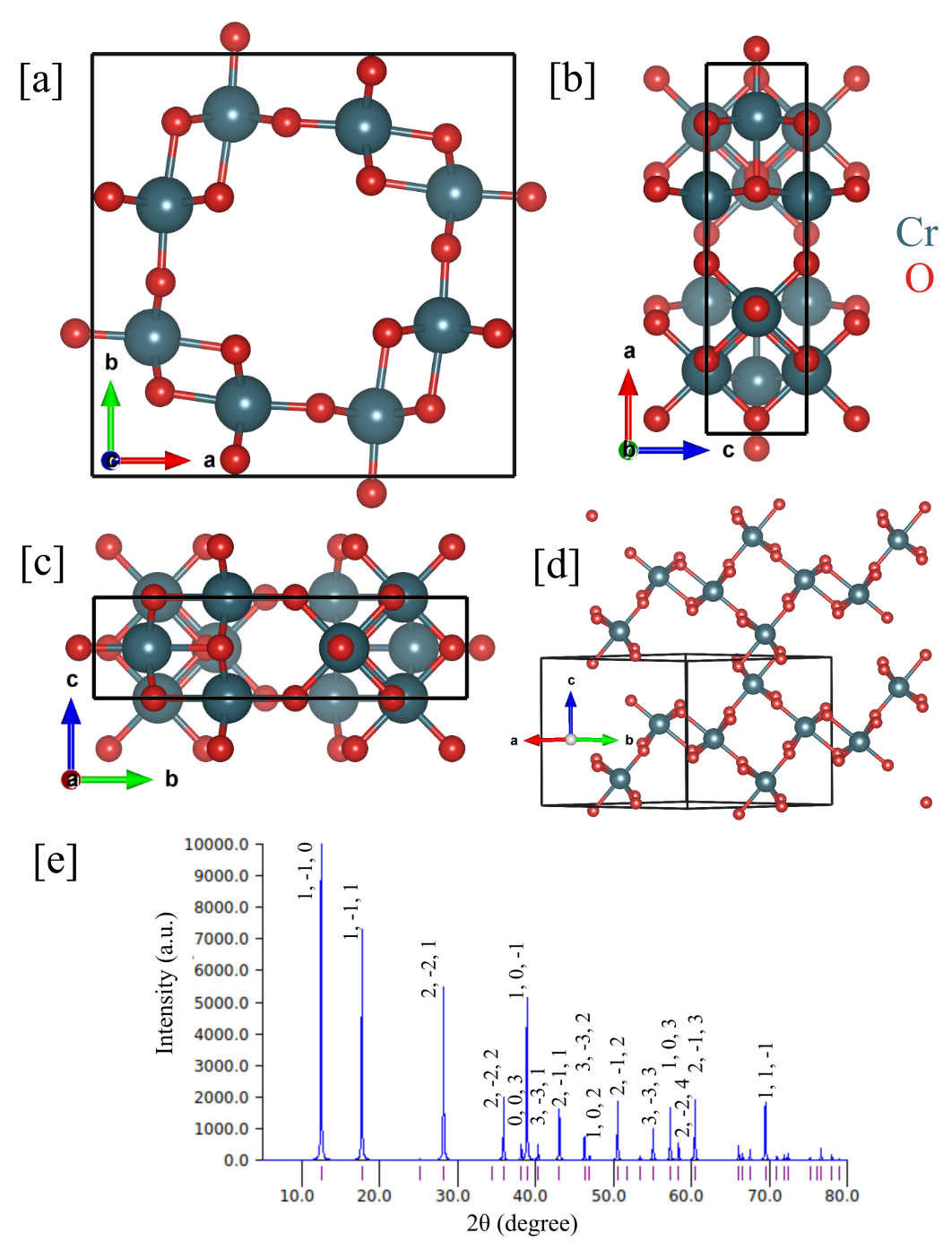}
	\caption{\label{fig-structure} (a-c) Conventional unit cell of h$-$\CrO ~crystal having I4/m symmetry, (d) Primitive unit cell, (e) Simulated X-ray Diffraction intensity in arbitrary unit with incident angle for h$-$\CrO. }
\end{figure}

\begin{figure*}[]
	\centering	
	\includegraphics[width=\textwidth]{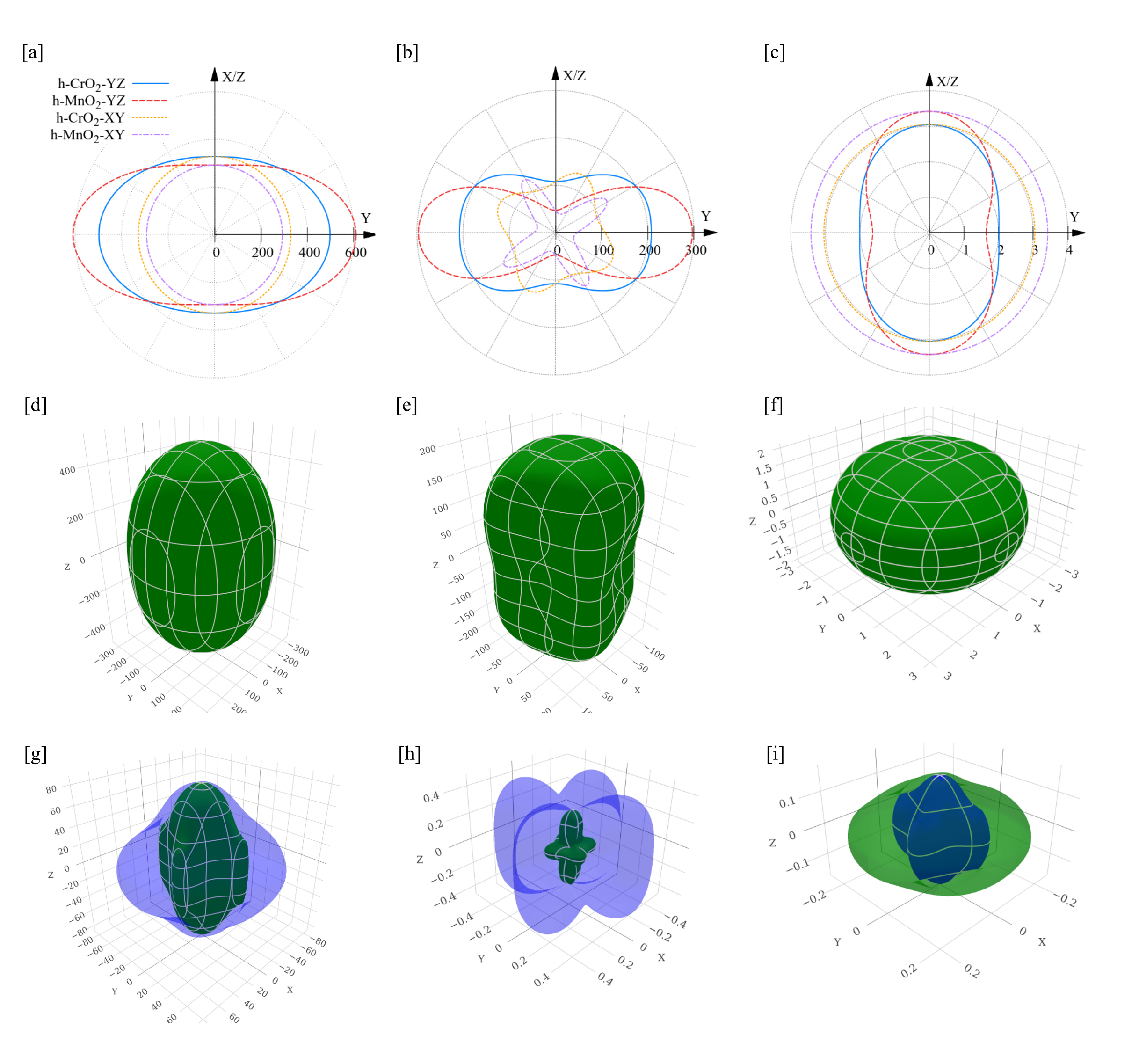}
	\caption{\label{fig-mechanical} (a) Bulk modulus (GPa), (b) Young modulus (GPa), (c) Compressibility (1/TPa) in XY and YZ plane of h$-$\CrO ~and h$-$MnO$_2$; Three dimensional plot of (d) Bulk modulus (GPa), (e) Young modulus (GPa), (f) Compressibility (1/TPa), (g) Shear modulus (GPa), (h) Poission ratio, (i) Inverse Pugh ratio (G/B$_H$)of h$-$\CrO. }
\end{figure*}
\section{Structure and Mechanical Properties}
{\par} The hollandite \cite{miura1986} CrO$_2$ follows body-centred tetragonal lattice having I4/m (82) crystal symmetry. The optimised lattice constants are found to be $a=b=9.992\AA$ and $c=2.702\AA$ calculated using PBE for spin unpolarised calculation. As depicted in Fig.\ref{fig-structure} (a$-$c), the Cr atoms coordinate with six neighbouring oxygen atoms forming edge-sharing CrO$_6$ octahedra. Such MO$_6$ type structure is also found in the rutile \CrO ~and is a common building block for many covalently bonded hard materials \cite{sun2019}. A $2\times2$ tunnel is formed in between the CrO$_6$ octahedras.  Including Hubbard terms $U=3.0$ and $J=0.87$ with PBE exchange-correlation the optimised lattice parameters are calculated as $a=b=9.880\AA$ and $c=2.978\AA$ \cite{korotin1998}. So, there is a relative underestimate of the volume by $7.23\%$ by the unpolarised calculation. Using same Hubbard parameters and PBEsol the lattice constants are found as $a=b=9.767\AA, ~c=2.928\AA$.

{\par} In Fig.\ref{fig-structure}(a-c), the conventional unit cell of h$-$\CrO~ crystal having eight formula units is shown from different angles. The primitive cell used for electronic structure calculations is presented in Fig.\ref{fig-structure}(d). The primitive cell contains four formula units. Now, for experimental identification of any crystal structure, XRD spectra is the key. Here in  Fig.\ref{fig-structure}(e), we provide the XRD spectra for Cu$_\alpha$ radiation. Reflection from (1,1,0), (1,1,1), (2,2,1) and (1,0,-1) crystallographic planes create the most prominent sharp peaks of XRD. This represents the signature character of the particular structure.

{\par} Experimentally two polymorphs of \CrO ~are prepared so far, the rutile type $\alpha-$\CrO (or r$-$\CrO) and orthorhombic $\beta-$\CrO (or o$-$\CrO). The lattice constants for both these structures predicted using PBEsol+U are agreeing with the experimental findings (see, Table$-$\ref{tab_mechanical}).

{\par} The hollandite structure of MnO$_2$ is known as $\alpha-$MnO$_2$. The calculated lattice constants for conventional unit cell using PBEsol+U are $a=b=9.787\AA$ and $c=2.903\AA$ match well with the reported experimental values $a=b=9.750\AA$ and $c=2.861\AA$ \cite{chen2012}.

\textbf{Elastic properties:~} Within elastic limit, according to Hook law, the stress ($\sigma_i$) and external strain ($e_j$) follow a linear relationship: 
$\sigma_i = \sum_{i,j=1}^{6} C_{ij} e_j$
, where, $C_{ij}$ is the elastic stiffness tensor. The orthorhombic system has nine independent components in the $6\times6$ matrix. The rutile structure being a part of type$-$I tetragonal system possesses six independent components. The hollandite structure which falls under type$-$II tetragonal class possesses seven independent components. Beside the stress-strain relationship, the elastic tensor can also be calculated from the total energy (E) using the harmonic approximation as: 
$C_{ij}=\frac{1}{V_0}\frac{\partial^2 E}{\partial e_i \partial e_j}$ 
, where, $V_0$ is the volume without any stress. The values of $C_{ij}$ are tabulated in Table \ref{tab_mechanical}.

\begin{table*}[]
	\centering
	\resizebox{\columnwidth}{!}{%
		\begin{tabular}{ccccccccccccccccccc}
			\hline
			\multirow{2}{*}{Material} & \multirow{2}{*}{Method} & \multirow{2}{*}{\begin{tabular}[c]{@{}c@{}}Lat.\\ Cnst.(\AA)\end{tabular}} & \multicolumn{10}{c}{Elastic Const. (GPa)} & \multirow{2}{*}{\begin{tabular}[c]{@{}c@{}}B$_H$\\ (GPa)\end{tabular}} & \multirow{2}{*}{\begin{tabular}[c]{@{}c@{}}B$_V$\\ (GPa)\end{tabular}} & \multirow{2}{*}{\begin{tabular}[c]{@{}c@{}}Y\\ (GPa)\end{tabular}} & \multirow{2}{*}{\begin{tabular}[c]{@{}c@{}}G\\ (GPa)\end{tabular}} & \multirow{2}{*}{$\nu$} & \multirow{2}{*}{$\frac{B_H}{G}$} \\
			&  &  & C$_{11}$ & C$_{22}$ & C$_{33}$ & C$_{44}$ & C$_{55}$ & C$_{66}$ & C$_{12}$ & C$_{13}$ & C$_{23}$ & C$_{16}$ &  &  &  &  &  &  \\ \hline
			r$-$CrO$_2$ & PBE+U & \begin{tabular}[c]{@{}c@{}}a=4.464\\ c=2.955\end{tabular} & 283.22 & - & 431.84 & 113.24 & - & 209.64 & 194.34 & 144.36 & - & - & 216.85 & 216.02 & 277.98 & 108.05 & 0.29 & 2.01 \\
			r$-$CrO$_2$ & PBEsol+U  & \begin{tabular}[c]{@{}c@{}}a=4.408\\ c=2.919\end{tabular} & 302.51 & - & 482.47 & 122.24 & - & 247.08 & 254.95 & 171.31 & - &  - & 252.39 & 245.13 & 268.17 & 101.36 & 0.32 & 2.49 \\
			r$-$CrO$_2$ & Expt. \cite{maddox2006}  & \begin{tabular}[c]{@{}c@{}}a=4.421\\ c=2.916\end{tabular}  & & & & & &  &  &  &  & & & 242 $\pm$ 2  & &  &   \\
			o$-$CrO$_2$ & PBE+U  & \begin{tabular}[c]{@{}c@{}}a=4.463\\b=4.465\\ c=2.956\end{tabular} & 272.66 & 272.60 & 419.28 & 116.02 & 115.98 & 210.44  & 197.75 & 151.11 & 150.98 & - & 216.67 & 193.94 & 267.48 & 103.33 & 0.29 & 2.10 \\	
			o$-$CrO$_2$ & PBEsol+U  & \begin{tabular}[c]{@{}c@{}}a=4.407\\b=4.406\\ c=2.920\end{tabular} & 306.76 & 305.99 & 460.08 & 131.78 & 131.83 & 238.70 & 229.74 & 169.98 & 169.76 & - &  244.49 & 216.75 & 295.88 & 113.95 & 0.30 &2.15  \\
			o$-$CrO$_2$ & Expt. \cite{maddox2006}  & \begin{tabular}[c]{@{}c@{}}a=4.425\\ b = 3.987\\ c= 2.683\end{tabular}  & & & & & &  &  &  &  & & & 181 $\pm$ 3  & &  & &  \\		
			h$-$CrO$_2$ & PBE+U & \begin{tabular}[c]{@{}c@{}}a=9.880\\ c=2.978\end{tabular} & 141.33 & - & 278.19 & 75.39 & -& 37.73 & 119.95 & 62.77 & - & 5.93 & 115.63 & 111.48 & 117.54 & 44.17 & 0.33 & 2.62 \\
			h$-$CrO$_2$ & PBEsol+U  & \begin{tabular}[c]{@{}c@{}}a=9.767\\ c=2.928\end{tabular} & 174.89 & - & 254.52 &  83.16 & - & 41.61 & 100.41 & 80.16 & - & -8.92 & 124.24 & 119.01 & 154.54 & 59.77 & 0.29 & 2.08 \\
			h$-$MnO$_2$ & PBE+U & \begin{tabular}[c]{@{}c@{}}a=9.920\\ c=2.935\end{tabular} & 139.56 & -& 309.64 &  77.66 & - & 37.75 & 100.58 & 58.91 & - & 12.32 & 111.01 & 120.98 & 131.12 & 50.31 & 0.30 & 2.21 \\
			h$-$MnO$_2$ & PBEsol+U & \begin{tabular}[c]{@{}c@{}}a=9.787\\ c=2.903\end{tabular} &  148.27 &  - & 331.03 &  82.05 & - & 38.25 & 112.11 & 66.75 & - & 13.81 & 121.11 & 128.58 & 131.93 & 50.03 & 0.32 & 2.42 \\ 
			h$-$MnO$_2$ & Expt.  \cite{chen2012} & \begin{tabular}[c]{@{}c@{}}a=9.750\\ c=2.861\end{tabular} & & & & & &  &  &  & &  &  &  & &  &  \\ \hline			
		\end{tabular}%
	}
	\caption{The lattice constants, elastic constants (C$_{ij}$), bulk moduli (B$_H$), Young moduli (Y), Sheer moduli (G), Poisson ratios ($\nu$), Pugh ratios (B$_H$/G) calculated from energy$-$strain relationship following the formula by Hill \cite{hill1952} and  bulk moduli calculated using equation of state proposed by Vinet et. al. \cite{vinet1986} (B$_V$) for rutile (r) and hollandite (h) type \CrO ~and h$-$\MnO. }
	\label{tab_mechanical}
\end{table*}

{\par} While all other structures are experimentally reported, h$-$\CrO~ is not been synthesised yet. Hence, a mechanical stability check is indispensable. Following the Born stability criteria extended to different crystal classes, the necessary and sufficient conditions for mechanical stability (Eq. \ref{eq_elastic}) are satisfied by all the structures are \cite{born1955,mouhat2014}: 
\small\begin{align}\label{eq_elastic}
\noindent
&\text{Orthorhombic: }C_{11}, C_{44}, C_{55}, C_{66} > 0; C_{11}C_{22} > C_{12}^2; \nonumber\\
&C_{11}C_{22}C_{33} + 2C_{12}C_{13}C_{23} - C_{11}C_{23}^2 - C_{22}C_{13}^2 - C_{33}C_{12}^2 > 0 \nonumber\\
&\text{Tetragonal$-$I: }C_{11}> |C_{12}|,~ C_{44}, C_{66} > 0; \nonumber \\
& 2C_{13}^2 < C_{33}(C_{11}+C_{12}) \nonumber\\
&\text{Tetragonal$-$II: }C_{11}> |C_{12}|;~ C_{44} > 0; \nonumber \\
& 2C_{13}^2 < C_{33}(C_{11}+C_{12});~ 2C_{16}^2 < C_{66}(C_{11}-C_{12})
\end{align}%

{\par} Beside the elastic stability of the proposed material, its dynamical stability test is necessary. As the vibrational modes (phonons) generate due to the relative kinetics of ions, for a dynamical stable structure, there should be no negative phonon mode. The phononic energy dispersion in Fig.\ref{fig-phonons}(a) delineate the dynamical stability. Chromium ions (Cr$^{4+}$) are much heavier than oxygen ions (O$^{-2}$), therefore, the lower energy phononic modes are populated by the contribution from Cr. Optical modes of higher frequency are mostly contributed by the kinetics of Oxygen ions.

{\par} The resistance against the external compression reflects in the bulk modulus of material. Using the Voigt-Reuss-Hill methodology the bulk moduli (B$_H$) are calculated \cite{voight1928,reuss1929,hill1952}. A set of separate calculation has been carried out to find the variation of energy with changing volume. The equation of state (EOS) fittings for those data provide another set of bulk moduli. In Table$-$\ref{tab_mechanical} B$_V$ represent the bulk moduli calculated using Vinet EOS \cite{vinet1987}. The calculated B$_V$ using PBEsol+U for r$-$\CrO~ is in line with the experimental value \cite{maddox2006}. From literature, the theoretically predicted values of the bulk modulus for r$-$\CrO ~are found as $261~GPa$ \cite{bendaoud2019}, $282~GPa$ \cite{wu2012}, $225~GPa$ \cite{li2012Structural}, $238~GPa$ \cite{alptekin2015}. Now, there is a mismatch of theoretical and experimental bulk moduli of o$-$\CrO ~structure (216.75 vs. 181). Maddox \etal have suspected that as the orthorhombic phase is not found in ambient pressure, the zero pressure volume can not be measured experimentally \cite{maddox2006}. Therefore, an effect of inadequate data may results in the EOS prediction for the phase.

{\par} Along with the Young moduli (Y), the sheer moduli (G), Poisson ratios ($\nu$), Pugh ratios (B$_H$/G) are calculated using the same Voigt-Reuss-Hill methodology and are tabulated in Table$-$\ref{tab_mechanical}. Relative to the orthorhombic and rutile structures, the hollandite structure has large vacuum present in between the CrO$_6$ polyhedra giving more space to accommodate the the deformation produced by external strains. As a result, B, Y and G values of of the h$-$\CrO~ phase are much less than other polymorphs. The $2\times2$ tunnel is also present in h$-$MnO$_2$, so, the values of elastic moduli come out to be almost similar to those ones of h$-$\CrO.

{\par} As for tetragonal class of materials the elastic constants follow the relation: $C_{12}=C_{21}$, hence, the variation of bulk moduli in the XY plane for both h$-$\CrO~ and h$-$MnO$_2$ are isotropic in nature, whereas, in ZX or ZY plane they show elliptical variation as depicted in Fig.\ref{fig-mechanical} (a). The eccentricity of the variation in ZY plane is lower for h$-$\CrO, while, the bulk modulus in XY plane is higher for the same. In contradiction to h$-$\CrO, YZ plot of bulk modulus of h$-$MnO$_2$ shows a dip at $Z=0$. The variation of compressibility in Fig.\ref{fig-mechanical}(c) is a confirmation of the fact that higher bulk modulus yields less compressibility. According to the Young moduli in Fig. \ref{fig-mechanical}(b), h$-$\CrO~ is less stiff along Z direction than h$-$MnO$_2$. The anisotropy of the elastic moduli can be visually confirmed through Fig.\ref{fig-mechanical}(d-e). In XY plane, the pattern of Young modulus shows it is hard to deform the shape of the $2\times2$ tunnel in h$-$\CrO. As h$-$MnO$_2$ is found in AFM ground state, its resistance against the displacement of Mn atoms is more robust.

\begin{figure}[b!]
	\centering	
	\includegraphics[width=\textwidth]{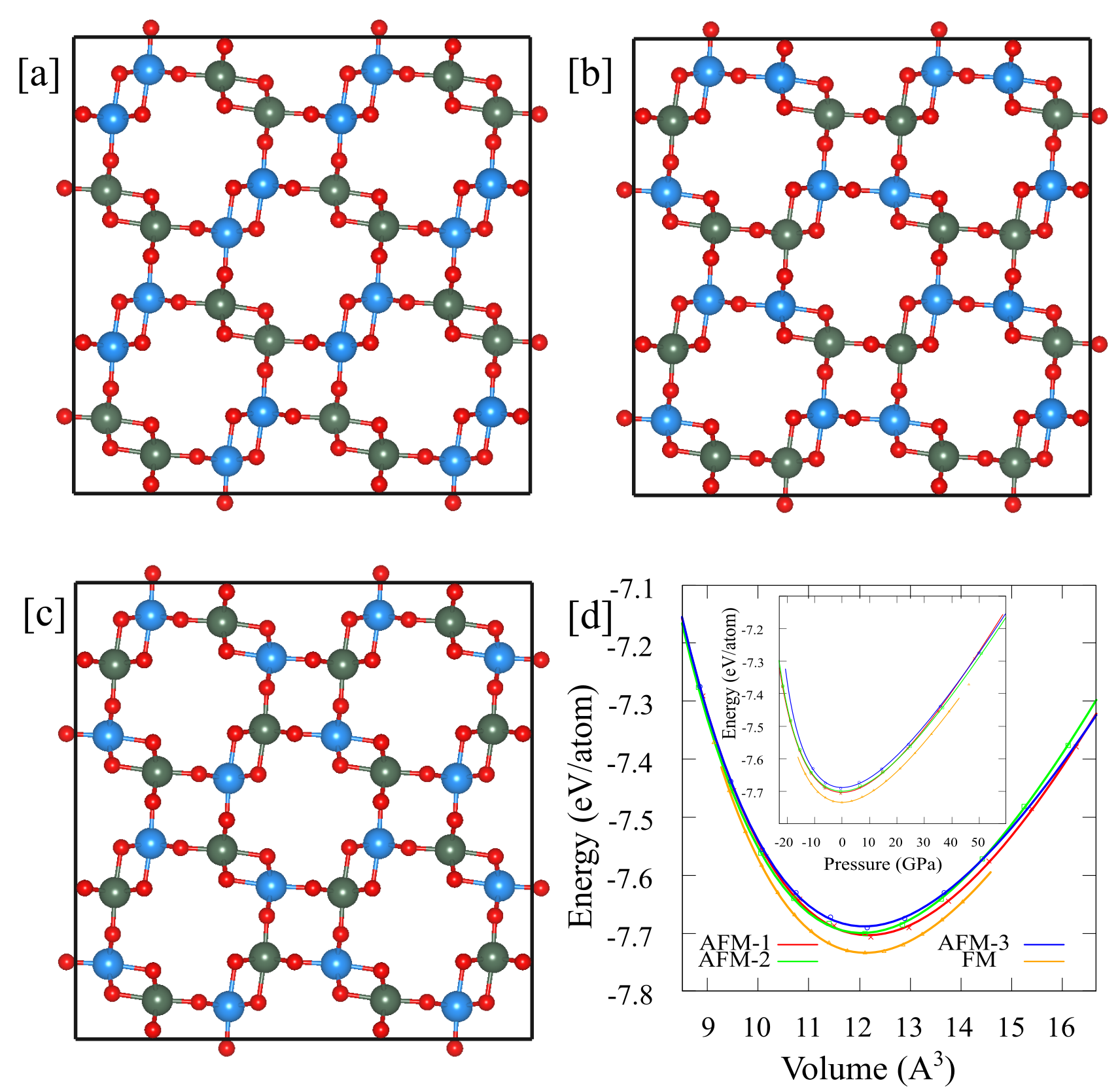}
	\caption{\label{fig_AFMconfig} (a-c) The anti-ferromagnetic spin configurations in a $2\times2\times1$ supercell of conventional unit cell. Green and blue dots represent atoms with opposite spins. (d) Energy vs. volume and energy vs. pressure (inset) plot for three anti-ferro and the ferromagnetic configurations. The solid lines are the Vinet equation of state fitted curves using the data shown as points \cite{vinet1986}.}
\end{figure}

\section{Electronic Structure}
{\par} The electronic structure and magnetic behaviour of \CrO~ has always been a curious case. While it is more likely for an TMO to be found in anti-ferromagnetic character, \CrO~ is found to be ferromagnetic in its ground state. For anti-ferro spin configuration we have taken three different arrangements as depicted in Fig.\ref{fig_AFMconfig}(a$-$c). The EOS plot confirms the ferromagnetic ground state of h$-$\CrO. The AFM$-1$ state is $30~meV/$atom higher in energy than the FM configuration. The AFM$-1$ and AFM$-2$ states are very close in energy, and for a range of external pressure it is showing a crossover. However, the ferromagnetic ground state remains stable under pressure, which indicates robustness of magnetic response.

{\par} Furthermore, the h$-$\CrO~ structure has been optimised using a variation of Hubbard term. Throughout the range of $U-J=0$ to $U-J=7$ the FM configuration remains the ground state as depicted in Fig.\ref{fig-E-U}. The energy variation with $U-J$ is non-linear with low eccentricity. For FM configuration this curvature is highest. 

{\par}On the other hand, h$-$MnO$_2$ which has similar structure is found to be in AFM$-3$ ground state with the spin distribution as in Fig.\ref{fig_AFMconfig}(c) \cite{paul2023,cockayne2012}. Interestingly, for h$-$\CrO~ this AFM$-3$ configuration possesses the highest energy. 

\begin{figure}[]
	\centering	
	\includegraphics[width=\textwidth]{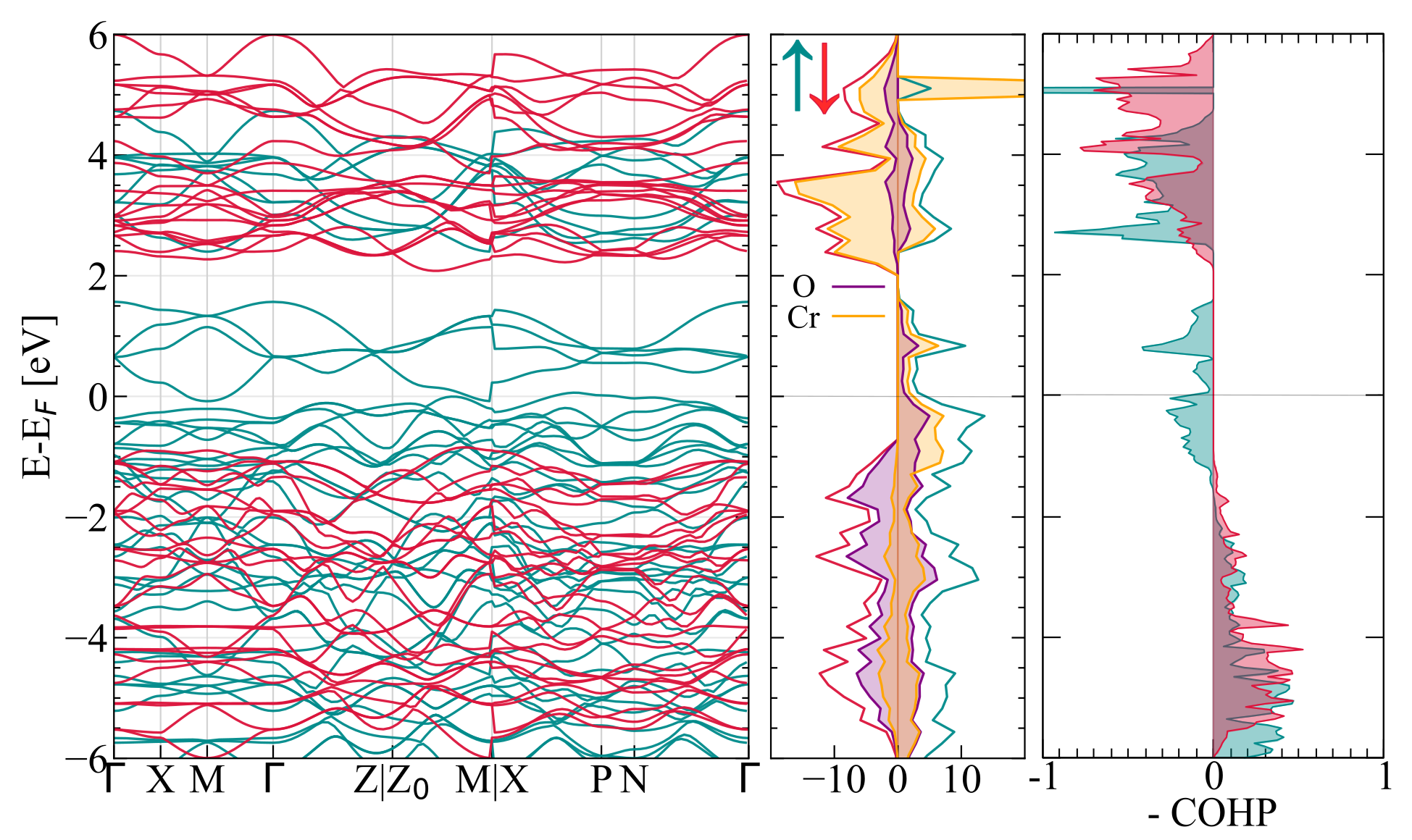}
	\caption{\label{fig-band-DOS} Energy band diagram, density of states per unit cell, and the negative crystal orbital Hamiltonian populations ($-$COHP) of h$-$\CrO ~crystal. }
\end{figure}

\begin{figure*}[]
	\centering	
	\includegraphics[width=\textwidth]{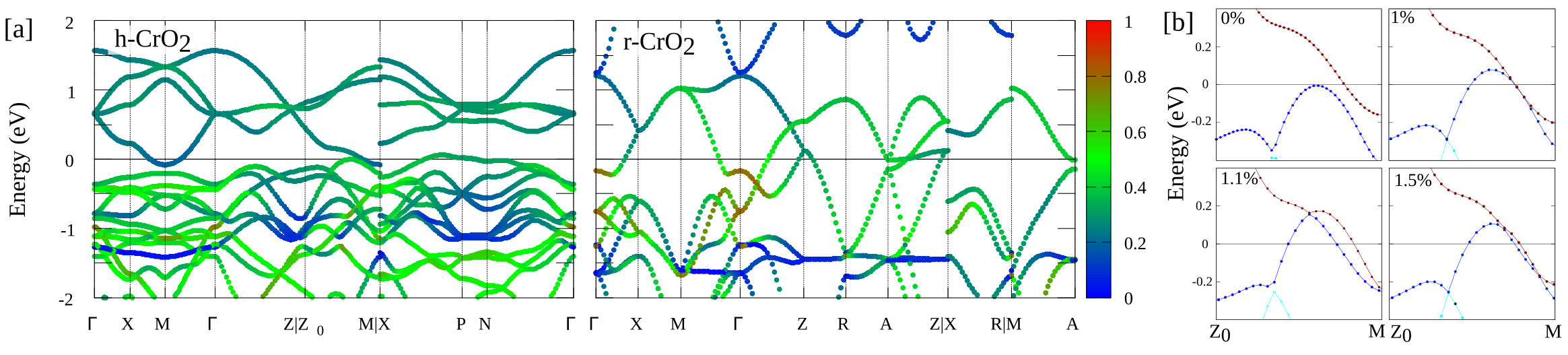}
	\caption{\label{fig-bands_pressure} (a) Percentage contribution from Oxygen orbitals for the bands near Fermi energy of h$-$\CrO~ and r$-$\CrO, (b) Evolution of bands crossing the Fermi energy for h$-$\CrO~ with uniaxial pressure along z-axis. }
\end{figure*}

{\par} The h$-$\CrO~ is a ferromagnetic material where the nature of bands for up and down spins are quite different. The metallic nature of one spin (up) is contradicted by the semiconductor nature for the other spin (down) channel making the h-\CrO~ a half-metal. The electronic band dispersion and the density of states (DOS) near the Fermi level (E$_F$) is depicted in Fig.\ref{fig-band-DOS}. The majority spin bands crosses E$_F$, while, the down spin channel shows a gap. The half-metallic bandgap is found to be $2.9~ eV$. There is no state over $-0.85~eV$ and below $2.07~eV$ relative to E$_F$ in minority spin channel. The half-metallic gap is indirect on the Z$_0-$M line of irreducible Brillouin Zone (BZ) edge (see, Fig.\ref{fig-phonons}). For majority spin channel, in conduction band, first four bands are separated from the others by $0.82~eV$ from $1.56~eV$ at $\Gamma$ to $2.38~eV$ at M w.r.t. E$_F$. An interesting feature is that the eigenvalues for all the bands at Z are equal with those at Z$_0$, so, the diagonally opposite points of upper surface of the irreducible BZ are equivalent.  

{\par} The DOS for majority spin channel is shifted lower than the minority spin channel. Shifting of DOS is an well-known sign of ferromagnetic character of material as well as the dissimilarity of DOS represents the ferromagnetic strength. For \CrO, the dissimilarity is vividly noticeable, so, the found magnetic moment of $2~\mu B$ per formula unit is quite justified.

{\par} Electronic bonding analysis can provide more insight of the ferromagnetic character of h$-$\CrO~ \cite{dronskowski2004}. For localised basis set the atomic orbital overlap is straight-forward to calculate, so, the overlap population weighted DOS (crystal orbital overlap population, COOP) can provide the information on the nature of bonding, anti-bonding, or non-bonding interaction. For DFT calculation involving plane wave basis, crystal orbital Hamiltonian populations (COHP) is a method that partitions band energies into pairwise atomic orbital interactions facilitating similar identification. There is no anti-bonding state below E$_F$ for down spin, whereas, for up spin, the anti-bonding state starts $-1.78~eV$. Such extended anti-bonding state is generated from Cr$-3d$ and O$-2p$ interaction which becomes clear from the orbital weighted bands in Fig.\ref{fig-pBAND}. The bonding anti-bonding nature is similar in rutile phase as well (see, Fig.\ref{fig-band-DOS-rutile}) 

{\par} To get a closer look on the orbital contribution on the full spaghetti of bands, the orbital weighted bands are plotted in Fig.\ref{fig-pBAND}. The lowest lying bands $40~eV$ below E$_F$ are originating from the the s and p orbitals of Cr, and, the majority spin bands are lower in energy than the minority spin bands. The O$-2s$ bands are also separated below $20~eV$. Near E$_F$, in valence band, the most contribution is coming from the O$-2p$ orbitals. For majority spin channel, the Cr$-3d_{xy}$ bands are separately visible. There is a small electron pocket at M  which is visible along X$-$M$-\Gamma$ and Z$_0-$M line. The pocket is conduced mostly from  Cr$-3d_{xy}$ and Cr$-3d_{x^2-y^2}$. Also, there is an hole pocket along X$-$P with most contribution coming from hybridised Cr$-3d_{xz}$ and O$-2p_x$. The band creating the hole pocket is more flat than the band responsible for electron pocket indicating heavier hole than electron at E$_F$.

{\par} From the partial DOS plot for Cr and O we have already noticed that the conduction bands are formed by Cr orbitals and the valence bands are mostly coming from O orbitals. Cr and O orbital mixing is better experienced for up spin. Now, r$-$\CrO~ is understood to be a negative charge transfer gap material \cite{korotin1998}. Korotin \etal have shown that a almost pure O$-2p$ band crosses the Fermi energy which acts as a electron/hole reservoir causing fractional occupation of Cr$-3d$ band at E$_F$. In Fig.\ref{fig-bands_pressure}(a) we have presented the oxygen weighted bands for both rutile and hollandite structure. In both cases, there is a Cr$-3d$ band below E$_F$ contributing mostly from $3d_{xy}$ orbital. However, in hollandite structure, it is not as much a separate unhybridised band, so, Cr$-3d_{xy}$ orbital in h$-$\CrO~ is not as localised  as in rutile counterpart. In r$-$\CrO, in the vicinity of BZ centre $\Gamma$, the band crossing E$_F$ holds its pure O$-2p$ character (brown dots), though hybridizes with Cr$-3d$ near Z. So, hybridisation is much anisotropic in r$-$\CrO~ for the said band. In h$-$\CrO~ there are two bands responsible for the metallic character, the band crossing E$_F$ around M has almost pure Cr$-3d$ character and the band crossing along X$-$P is a hybridised Cr$-d_{xz}$ and O$-2p_x$ band. In both polymorphs, the band mixing between Cr$-3d$ and O$-2p$ is accountable for the half-metallic ferromagnetic character, though, the nature of hybridisation is different. 



\begin{figure}[]
	\centering	
	\includegraphics[width=\textwidth]{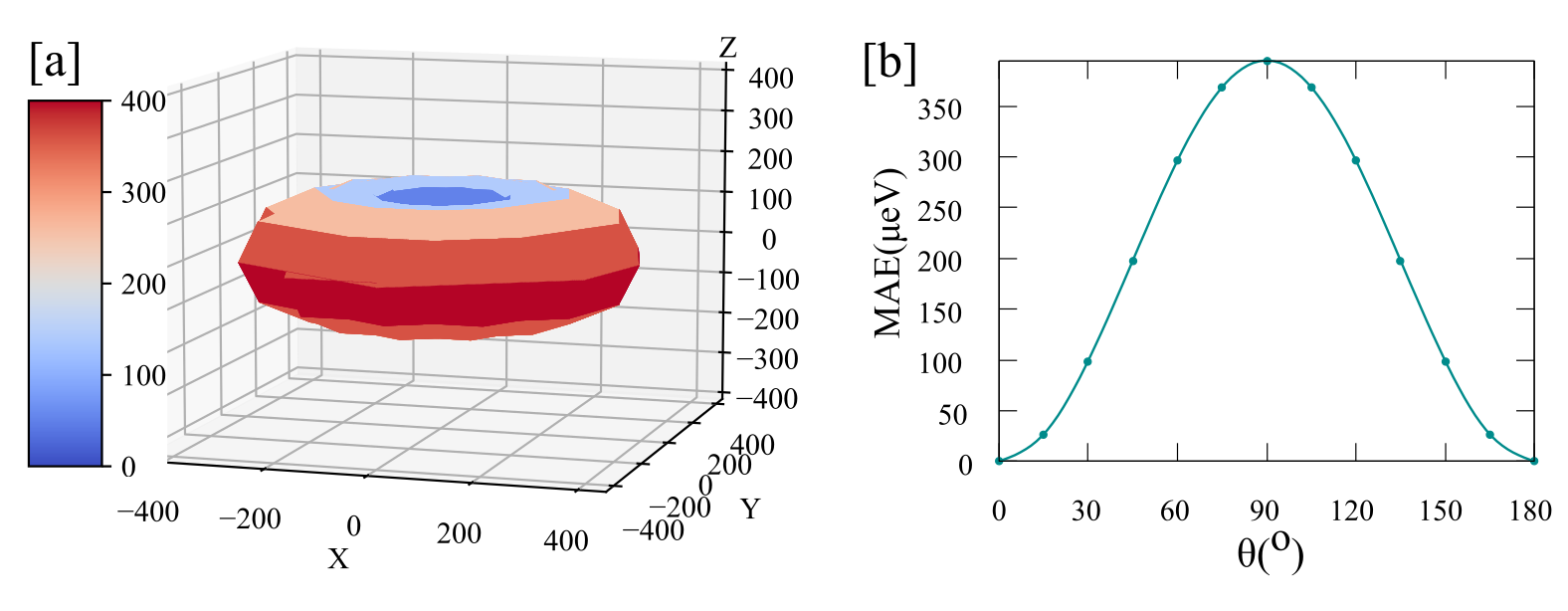}
	\caption{\label{fig-MAE} Magneto-crystalline anisotropic energy ($\mu eV$) variation of h$-$\CrO ~crystal in (a) 3$-$dimension, and in (b) X$-$Y plane, where $\theta$ is the angle with Z-axis. }
\end{figure}

\textbf{Magneto-crystalline anisotropy:~} h$-$\CrO~ possesses a rare porous hollandite crystal structure and belongs to a scarce ferromagnetic TMO family. It would be interesting to see how the crystal structure tunes the direction of magnetic moment in this system. Magneto-crystalline anisotropy is a phenomenon manifested by the variation of internal energy depending on the direction of magnetization in any material. As the orbit is strongly coupled to the crystal structure (lattice), so, changing the orientation of spin is resisted through spin-orbit coupling giving rise to MAE \cite{daalderop1990, sander2004}. The spatial variation of MAE is presented in Fig.\ref{fig-MAE}. The three dimensional plot is showing an isotropy in X$-$Y plane (azimuthal independence). Here to mention that the bulk modulus for this structure has also shown azimuthal symmetry (refer Fig.\ref{fig-mechanical}(d)). This is a feature of its layered structure. However, tilting the spin through making an angle $\theta$ with Z-axis (easy axis) demands external energy. The MAE variation with $\theta$ shows that the hardest axis lies on X$-$Y plane with highest value of $MAE=394.96~\mu Ev$. 

\textbf{Topological character under uniaxial pressure:~}Ferromagnetic Weyl materials poised to be integral part of rich Physics \cite{kanagaraj2022,wang2018large}. Occurrence of Weyl points in r$-$\CrO~ has been reported recently \cite{wang2018}. As we see in Fig.\ref{fig-bands_pressure}, along Z$_0-$M near E$_F$ two bands are almost touching each other within $\pm 0.5~eV$ range about E$_F$. Referring to Fig.\ref{fig-pBAND}, we can observe that one of these bands have strong Cr$-3d_{xz}$ contribution and, another with O$-2p_z$ orbital share. This makes us curious if uniaxial pressure along z-axis can bring some fundamental change of band topology (see, Fig.\ref{fig-bands_pressure}(b)). With 1$\%$ pressure, the bands touches each other. With higher pressure, band inversion feature is observed maintaining the band-topology conserved always. These two bands come close at two points, one above E$_F$,  another below E$_F$. This is a signature of type$-$II Weyl material \cite{yan2017topological}. More comprehensive study on the topological aspect of this material is aimed to be explored in future.

\section{Conclusion}
{\par}$\alpha-$MnO$_2$ structure is one of its kind with a $2\times2$ tunnel to accommodate foreign elements which can be found to be useful in different applications. We have predicted another TMO having a similar structure, yet drastically different electronic and magnetic character. Even, it demonstrates interesting mechanical and electronic behaviour compared with other Chromium dioxides. A detailed side-by-side study reveals the underlying Physics of such a vibrant character.

{\par} The system is mechanically and dynamically stable. It exhibits anisotropic elastic moduli and is less stiff than the similar h$-$MnO$_2$ structure. The lattice parameters and elastic constants calculated for hollandite h$-$MnO$_2$ and rutile r$-$\CrO~ are at par with the experimental values exhibiting the reliability of methodology utilised for the proposed h$-$\CrO~ crystal. 

{\par}The h$-$\CrO~ is a half-metallic ferromagnet having a half-metallic bandgap of $2.9~eV$. The ferromagnetic nature is a result of strong hybridisation of Cr$-3d$ and O$-2p$ electrons. The band crossing is minimal with only one electron pocket around M conduced mostly from  Cr$-3d_{xy}$ and Cr$-3d_{x^2-y^2}$, and one hole pockets along X$-$P with most contribution coming from hybridised Cr$-3d_{xz}$ and O$-2p_x$ at Fermi level. The occurrence of electron-hole pocket is at different k$-$points, so, the direct electron-hole coupling is not possible. However, the prospect of phonon mediated superconductivity is yet to be investigated.

{\par}The crystal shows ample magneto-crystalline anisotropy with the easy axis perpendicular to the plane of the structure (Z$-$axis) with highest value of  $MAE=394.96~\mu Ev$, while the azimuthal angle independence of MAE is evident. Though there are two bands almost touching together near Fermi level, the band topology remains conserved upon uniaxial pressure. However, the transformation of bands near Fermi energy with minimal uniaxial pressure may pave the way for further detailed study on the topological aspects of this material.  

{\par} To sum up, we can conclude that a new stable transition metal oxide with vibrant physical properties is proposed which may further ignite the interest of Physics community.

\section*{Acknowledgement}
The author want to thank Prof. Debnarayan Jana of  University of Calcutta, India for enriching discussion and Prof. Sajeev John of University of Toronto, Canada for providing computational facility.

\bibliographystyle{elsarticle-num}
%

\appendix*
\renewcommand\thefigure{A.\arabic{figure}} 
\section{~}

\setcounter{figure}{0} 
\begin{figure}[h]
	\centering	
	\includegraphics[width=\textwidth]{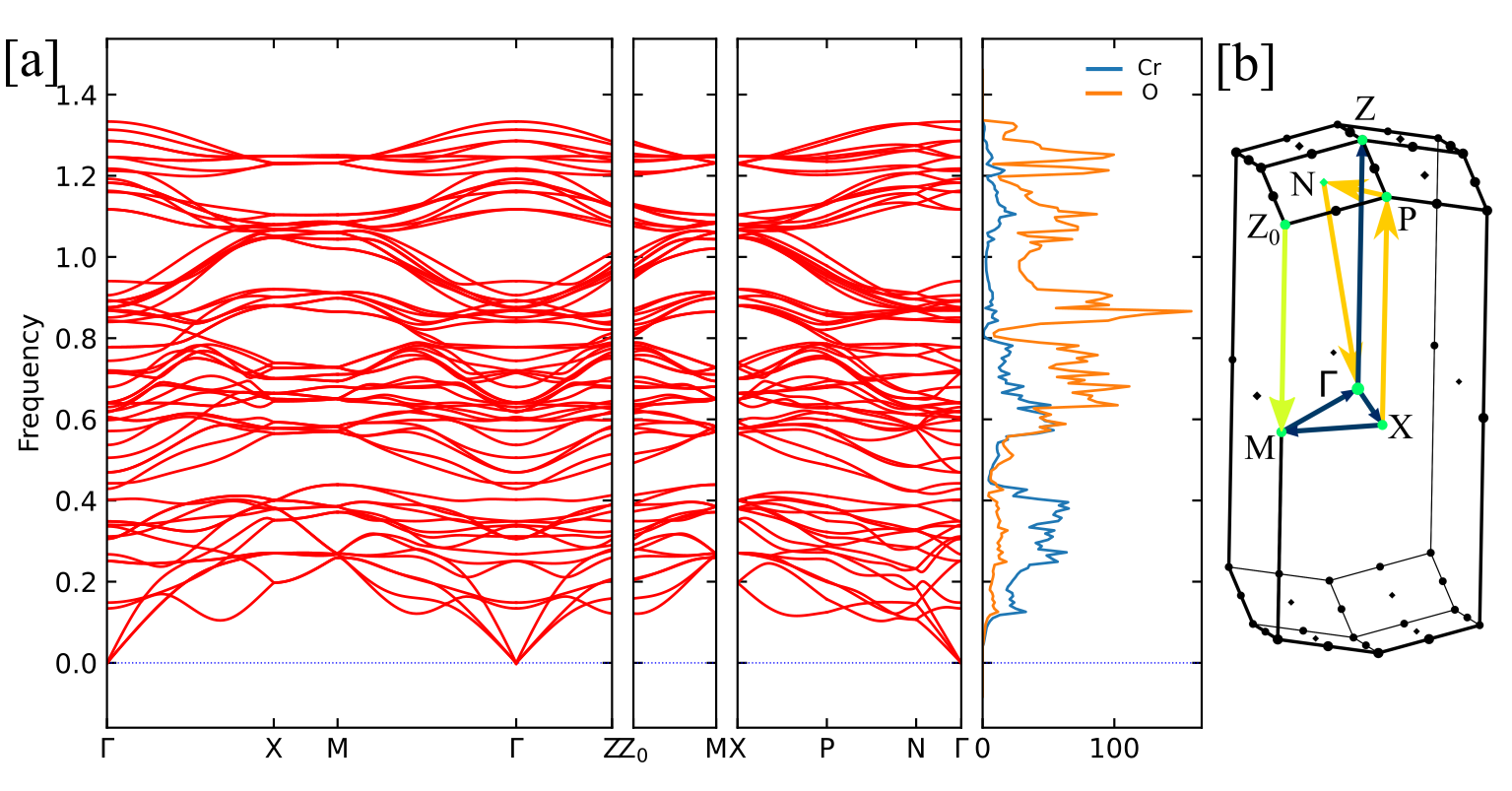}
	\caption{\label{fig-phonons} (a) Phononic band structure and DOS of h$-$\CrO; (b) The irreducible Brillouin zone and the k-path selected for band structure. }
\end{figure}

\begin{figure}[]
	\centering	
	\includegraphics[width=0.8\textwidth]{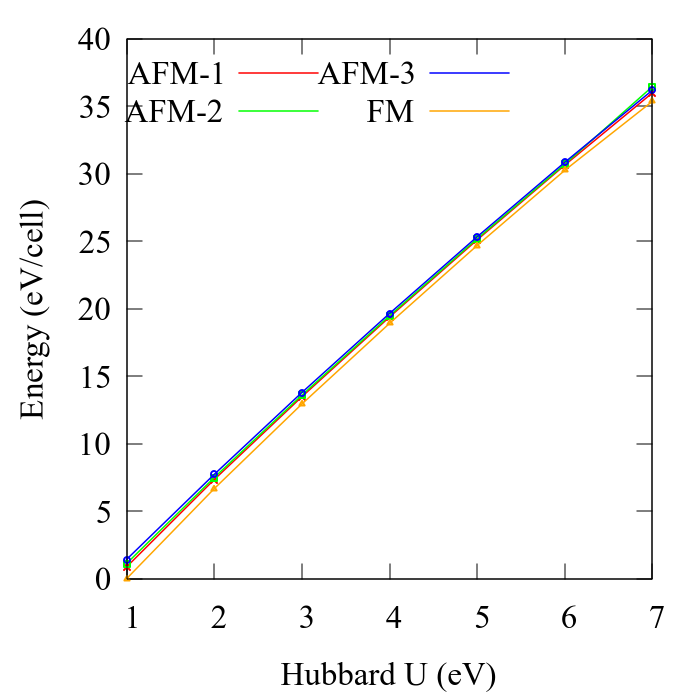}
	\caption{\label{fig-E-U} Total energy w.r.t. ground state energy per unit cell for different magnetic configurations of h$-$\CrO. }
\end{figure}

\begin{figure}[]
	\centering	
	\includegraphics[width=\textwidth]{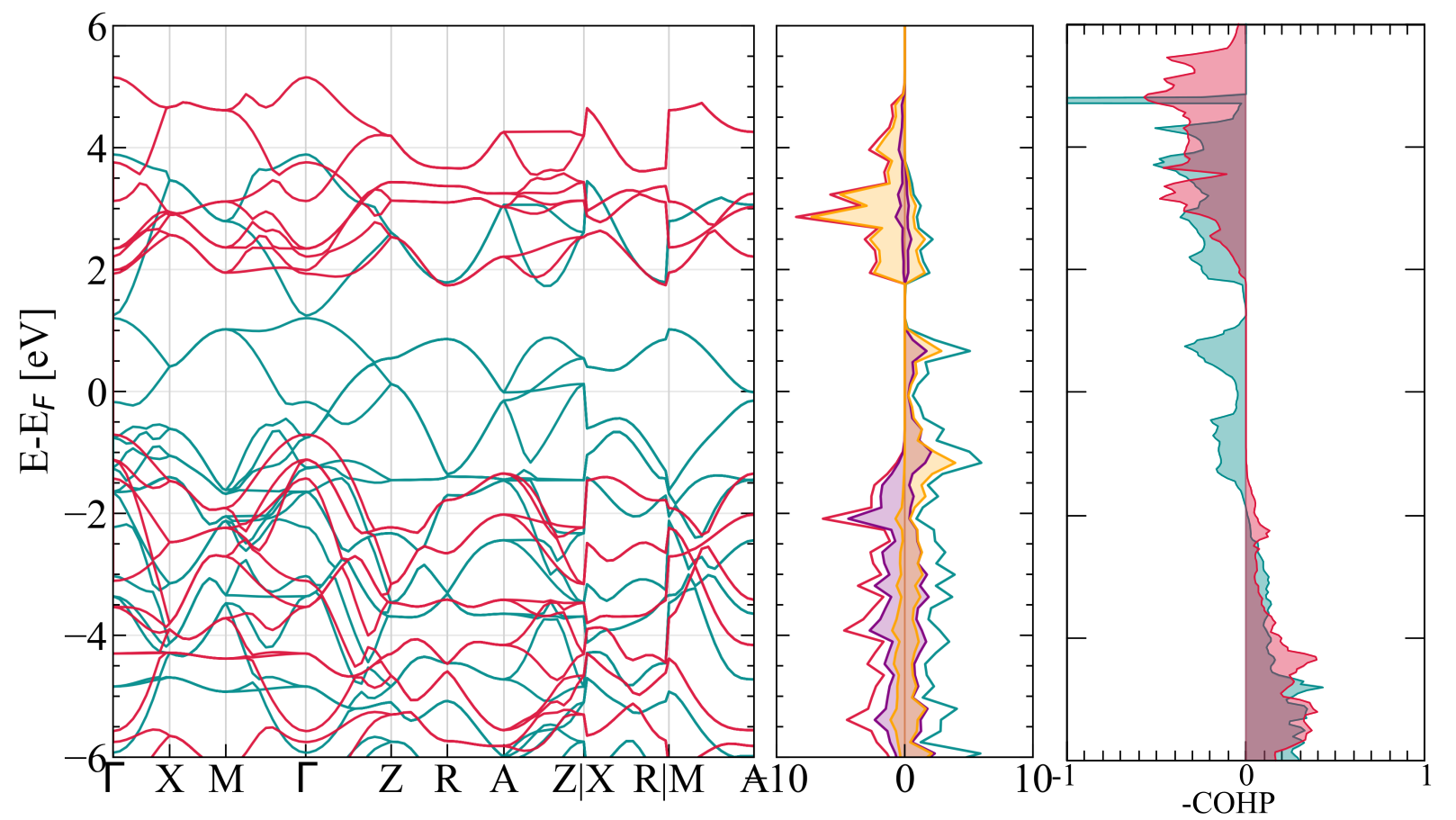}
	\caption{\label{fig-band-DOS-rutile} Energy band diagram, density of states, and the negative crystal orbital Hamiltonian populations ($-$COHP) of r$-$\CrO ~crystal. }
\end{figure}

\begin{figure*}[]
	\centering	
	\includegraphics[width=0.7\textwidth]{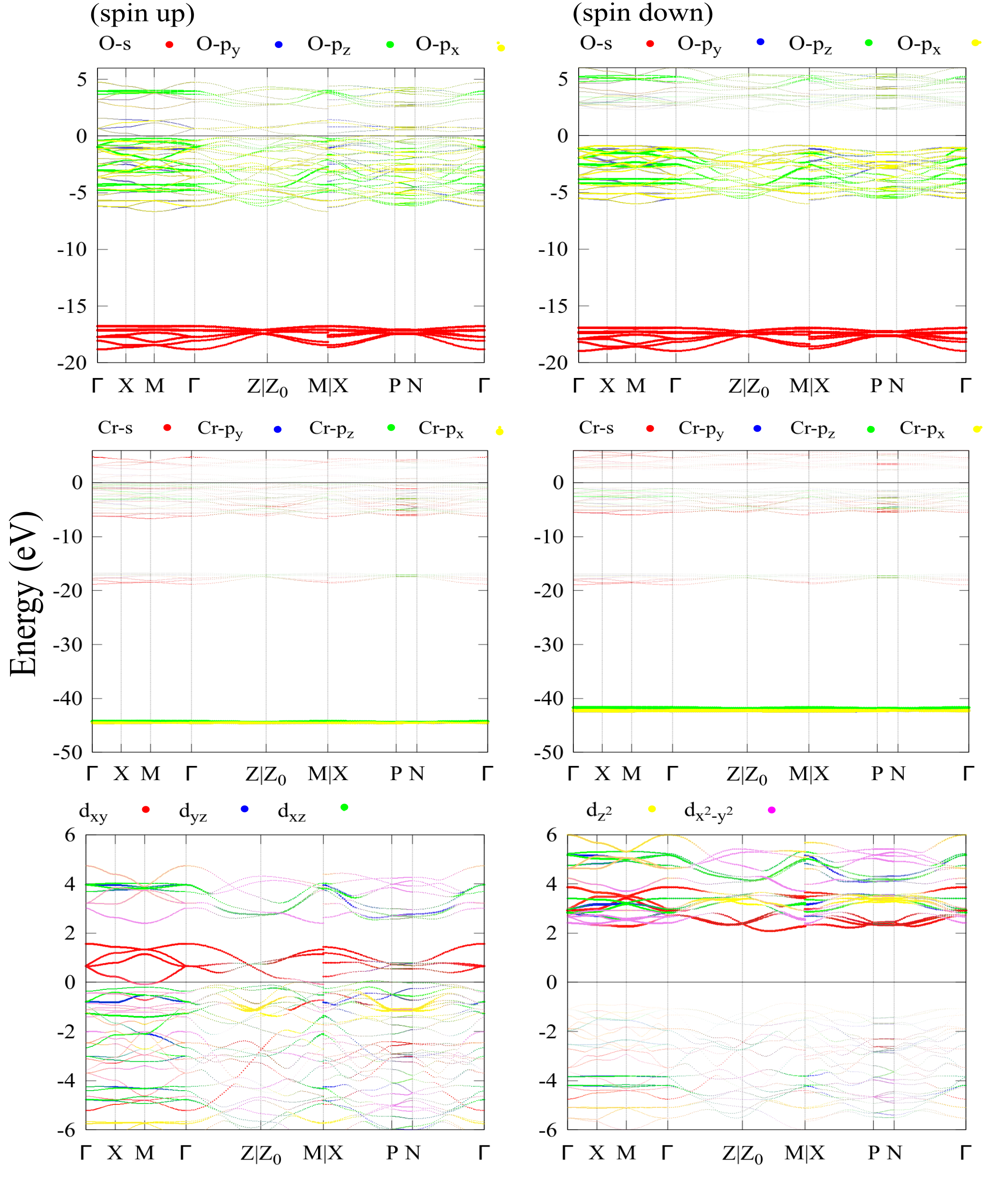}
	\caption{\label{fig-pBAND} Orbital weighted band dispersion of h$-$\CrO. }
\end{figure*}

\end{document}